УДК 378:147:51:004

**Шишкіна Марія Павлівна**
доктор педагогічних наук, старший науковий співробітник
Інститут інформаційних технологій і засобів навчання НАПН України, м. Київ, Україна
*shyshkina@iitlt.gov.ua*

**Когут Уляна Петрівна**
кандидат педагогічних наук, доцент кафедри інформатики та обчислювальної математики
Дрогобицький державний педагогічний університет імені Івана Франка, м. Дрогобич, Україна
*ulyana_kogut@mail.ru*

# ВИКОРИСТАННЯ ХМАРО ОРІЄНТОВАНОГО КОМПОНЕНТУ НА БАЗІ СИСТЕМИ MAXIMA У ПРОЦЕСІ НАВЧАННЯ ДОСЛІДЖЕННЯ ОПЕРАЦІЙ

**Анотація.** У статті досліджено проблеми використання систем комп'ютерної математики (СКМ) у сучасному високотехнологічному середовищі, зокрема, висвітлено перспективні шляхи запровадження хмаро орієнтованих компонентів на базі СКМ, що є суттєвим чинником розширення доступу до них як засобу навчальної і дослідницької діяльності у галузі інформатичних і математичних дисциплін. Визначено роль СКМ у підготовці бакалаврів з інформатики й особливості їх педагогічного застосування у навчанні дослідження операцій. Розглянуто основні характеристики СКМ MAXIMA та шляхи організації доступу до неї як у локальній, так і у хмаро орієнтованій реалізації. Наведено результати експертного оцінювання хмаро орієнтованого компонента на базі системи MAXIMA у навчальний процес дослідження операцій.

**Ключові слова**: хмарні технології; освітнє середовище; вищий навчальний заклад; гібридна сервісна модель; електронні освітні ресурси.

## 1. ВСТУП

**Актуальність дослідження.** У сучасному інформаційно-освітньому середовищі виникають нові моделі організації навчальної діяльності, що ґрунтуються на інноваційних технологічних рішеннях щодо організації інфраструктури середовища, до числа яких належать хмаро орієнтовані.

Питання налаштування інформаційно-технологічної інфраструктури навчального закладу на потреби користувачів, організація засобів і сервісів цього середовища так, щоб можна було максимальною мірою реалізувати педагогічний потенціал використання сучасних ІКТ, досягнути підвищення рівня результатів навчання, а також поліпшення організації процесів науково-педагогічної діяльності, передбачають обґрунтування шляхів організації доступу до такого різновиду програмного забезпечення, як системи комп'ютерної математики (СКМ), що є провідним типом засобів навчання математичних та інформатичних дисциплін.

Навчання дослідження операцій у системі підготовки фахівців з інформатики у педагогічному університеті відіграє особливу важливу роль, бо поєднує в собі як фундаментальні поняття і принципи різних математичних та інформатичних дисциплін, так і прикладні моделі й алгоритми їх застосування. У ній реалізуються основні наукові підходи до математичного моделювання процесів, обґрунтування рішень, математичного опису базових понять і принципів реалізації інформаційних процесів опрацювання даних, що власне і є предметом комп'ютерного моделювання в інформатиці

Використання СКМ у процесі навчання дослідження операцій дозволить: змінити акценти у доборі теоретичного матеріалу; збільшити частку задач на побудову математичних моделей реальних оптимізаційних задач і їх дослідження за допомогою СКМ; запровадити завдання на порівняння результатів, одержаних за допомогою чисельних методів оптимізації, описаних однією з мов програмування, і за допомогою вбудованих засобів СКМ, і їх аналіз для різних вхідних даних, а також завдання на програмування в середовищах математичних пакетів чисельних методів оптимізації та їх дослідження.

**Постановка проблеми.** Аналіз вітчизняного і міжнародного досвіду використання ІКТ у процесі навчання інформатичних дисциплін свідчить, що такий клас програмних засобів навчального призначення, як системи комп'ютерної математики (СКМ) постійно привертає увагу дослідників [8; 9; 10; 14]. Ці системи, що є комплексними, багатофункціональними, досить потужними і в той же час простими у використанні, стають незамінними у підтримуванні різноманітних процесів чисельних обчислень, візуалізації закономірностей, реалізації символьних операцій, алгоритмів і процедур [12; 14]. СКМ є середовищем для проектування та використання програмних засобів підтримування навчання інформатичних дисциплін, формуючи інноваційні педагогічні технології [12].

В останні роки засоби і технології навчання інформатичних дисциплін отримали подальший розвиток, зокрема, на основі концепції хмарних обчислень. Ця концепція суттєво змінює існуючі уявлення щодо організації доступу та інтеграції додатків, тому виникає можливість управління більш великими ІКТ-інфраструктурами, що дозволяють створювати і використовувати незалежно один від одного як індивідуальні, так і колективні «хмари» в межах загального хмаро орієнтованого освітнього простору [4; 12].

Локалізація таких засобів навчального призначення, як СКМ «у хмарі» є перспективним напрямом їх розвитку, коли виникає більше можливостей адаптації середовища навчання до рівня навчальних досягнень, індивідуальних потреб і цілей того, хто вчиться. Відбувається розширення «спектру» дослідницької діяльності як за рахунок фундаменталізації змісту навчання інформатичних дисциплін, так і за рахунок розширення доступу до засобів дослідницької діяльності. У зв'язку з цим потребують уваги питання обґрунтування теоретичних і науково-методичних засад формування і використання хмаро орієнтованих компонентів навчального призначення на базі систем комп'ютерної математики, визначення переваг і недоліків різних підходів до їх розгортання, дослідження й аналіз досвіду їх упровадження.

Використання СКМ Maxima у процесі навчання дослідження операцій спрямоване на формування ІКТ-компетентностей студентів, зокрема і у хмаро орієнтованому середовищі, завдяки: ознайомленню з функціональними характеристиками СКМ Maxima; виробленню навичок математичного дослідження прикладних задач, зокрема побудови математичних моделей; освоєнню спеціальної термінології; опанування програмування в СКМ Maxima; здобуття необхідної бази знань для вивчення інших дисциплін; підвищення рівня інформатичної підготовки шляхом широкого використання СКМ і хмаро орієнтованих систем у навчальному процесі та науково-дослідній роботі.

**Аналіз стану дослідження проблеми.** Формування хмаро орієнтованого середовища є перспективним напрямом інформатизації освіти, що визнаний пріоритетним міжнародною освітньою спільнотою [2, 17, 24], інтенсивно розробляється нині у різних галузях освіти, зокрема математичної та інженерної [27, 28, 29]. Тенденції впровадження у навчальних закладах хмарних технологій для організації доступу до програмного забезпечення, що застосовується для різних видів колективної роботи, під час здійснення наукової і навчальної діяльності, дослідно-конструкторських розробок, реалізації проектів, обміну досвідом тощо набувають особливої актуальності в останній час [3, 15, 27, 28].

У контексті використання програмного забезпечення навчального призначення у хмаро орієнтованому освітньому середовищі слід зазначити досвід Массачусетського технологічного інституту (МІТ) щодо розгортання хмарного доступу до математичних пакетів прикладних програм, зокрема – *Mathematica, Mathlab, Maple, R, Maxima* [29]. Проектування хмарних додатків актуально не лише для підтримування навчання математичних дисциплін, де це обумовлено потребою використання потужних серверів для виконання обчислень, а також і для багатьох інших галузей, зокрема, організації лабораторій віддаленого доступу, комп'ютерного дизайну та навчання [27, 28].

Предметом сучасних досліджень постає випробування різних моделей доступу до програмного забезпечення навчального призначення, зокрема, засобами віртуальної машини [28]; порівняльний аналіз програмного забезпечення з точки зору педагогічного використання, встановленого «у хмарі», визначення чинників найдоцільнішої організації освітнього середовища навчального закладу [15].

Методичні особливості навчання методів оптимізації та дослідження операцій з використанням WEB-СКМ проаналізовано у роботі Триуса Ю. В. [11]. Детально описані засоби графічного середовища системи комп'ютерної математики Maxima для моделювання анімацій, наведені приклади створення моделей анімаційних наочностей і їх використання для розвитку навчально-дослідницьких умінь, Бугаєць Н.О .[5].

Суттєвою особливістю хмарних обчислень є можливість динамічного постачання обчислювальних ресурсів та програмно-апаратного забезпечення, його гнучким налаштуванням на потреби користувача. За цього підходу організується доступ до різних типів програмного забезпечення навчального призначення, що може бути як спеціально встановлено на хмарному сервері, так і надаватися як загальнодоступний сервіс (знаходитися на будь-яких інших носіях електронних даних, що є доступні через Інтернет) [3, 14, 20, 28]. Через це потребує вивчення питання: які виникають способи і моделі педагогічної діяльності, як змінюється роль електронних освітніх ресурсів (ЕОР) і підходи до їх проектування, які засоби, моделі і шляхи організації доступу до них доцільно впроваджувати з огляду на існуючі тенденції формування у навчальних закладах хмаро орієнтованого середовища.

**Мета статті**: обґрунтування доцільності створення і використання хмаро орієнтованого компоненту на базі системи Maxima у процесі навчання дослідження операцій у педагогічному навчальному закладі, визначення перспективних шляхів його запровадження.

## 2. МЕТОДИ ДОСЛІДЖЕННЯ

Дослідження спирається на методи теоретичного аналізу, систематизації й узагальнення наукових фактів про педагогічні процеси і явища, методи системного аналізу і моделювання, педагогічні спостереження й узагальнення педагогічного досвіду, а також результати педагогічного експерименту. Дослідження здійснювалося у межах виконання планових НДР, що виконувалися в Інституті інформаційних технологій і засобів навчання НАПН України і на кафедрі інформатики та обчислювальної математики Дрогобицького державного педагогічного університету імені Івана Франка, у ході чого було обґрунтовано і проаналізовано науково-методичні засади використання хмаро орієнтованого компоненту на базі системи Maxima у процесі навчання інформатичних дисциплін фахівців з інформатики.

В інформатиці використовують такі міжпредметні методи та процедури, як аналіз і синтез, індукція і дедукція, візуалізація та формалізація, алгоритмізація і програмування, інформаційно-логічне, математичне та комп'ютерне моделювання, програмне управління, експертне оцінювання, ідентифікація та інші. Їх треба опановувати комплексно, інакше не відбувається достатнього рівня оволодіння матеріалом інформатичних дисциплін.

## 3. РЕЗУЛЬТАТИ ДОСЛІДЖЕННЯ

### 3.1. Особливості використання системи Maxima у процесі навчання інформатичних дисциплін

В умовах формування інформаційного суспільства зростає роль підготовки висококваліфікованих фахівців, які здатні до продуктивної діяльності в цьому суспільстві. Тому необхідним є пошук нових методичних підходів до організації навчання, що сприяли б глибокому засвоєнню і розумінню базових понять, правил, принципів і методів навчання дисциплін, їх взаємозв'язку із суміжними дисциплінами, а також шляхів їх використання на практиці. Перспективним напрямом видається інтегрування у процес навчання дослідження операцій систем комп'ютерної математики, за допомогою яких можна, з одного боку, автоматизувати деякі рутинні дії, зосередивши увагу студента на опануванні понять і принципів, що вивчаються, а з іншого боку, виявити міжпредметні зв'язки різних дисциплін, дослідивши, як ті чи інші фундаментальні поняття реалізуються у прикладних галузях.

Використання хмаро орієнтованих засобів проектування СКМ є суттєвим чинником розширення доступу до них у процесі навчальної і дослідницької діяльності у галузі інформатичних та математичних дисциплін. Якщо у випадку застосовувалися лише локальної версії засобу дослідницька діяльність з ним відбувається тільки у спеціально створених навчальних ситуаціях, то з використанням хмаро-орієнтованої версії більше уваги можна приділити самостійній роботі, і дослідницька діяльність поширюється і у поза-аудиторний час [12].

Застосування математичних пакетів до розв'язування практичних задач передбачає розуміння проблематики навчальної дисципліни для правильного використання СКМ; розуміння методології розробки алгоритму від математичної ідеї до формулювання алгоритму та вміння застосувати цю методологію; вміння здійснювати обгрунтування й оцінювання складності алгоритму за часом виконання і необхідної пам'яті [14, с. 138].

Для наукових цілей вибір СКМ залежить від вхідних даних та результату, що необхідно отримати. Наприклад, фізику-теоретику більш цікава аналітична модель досліджуваного явища чи об'єкта, тому доцільніше використовувати пакети, такі як Mathematica, Maple, Maxima. Фізикам-експериментаторам для опрацювання великих масивів даних зручно використовувати систему MATLAB [14, с. 138].

Особливу увагу звернемо на систему Maxima, оскільки вона є легка в опануванні, не поступається у розв'язуванні задач таким системам як Maple та Mahtematica та є вільно поширюваною. Вона оснащена системою меню, що дає змогу виконувати символьні перетворення, розв'язувати рівняння, обчислювати границі, похідні, інтеграли тощо, не знаючи мови для опису команд щодо виконання цих дій. Тому систему Maxima можна використовувати для вивчення інформатичних і математичних дисциплін навіть на першому курсі педагогічного університету [14]. Застосування системи Maxima не викличе ніяких труднощів у студентів під час розв'язування задач математичного аналізу та лінійної алгебри – від студентів вимагається тільки правильно вибрати пункт меню та ввести вираз. Проте для програмування у системі Maxima потрібні знання мови та синтаксису, а також і певних команд [14, с. 138].

Навчання майбутніх фахівців з інформатики дослідження операцій потребує особливої уваги, бо поєднує в собі як фундаментальні поняття і принципи різних математичних та інформатичних дисциплін, так і прикладні моделі та алгоритми їх застосування.

Метою використання СКМ у процесі підготовки майбутніх фахівців з інформатики є формування в них здатності до успішного використання інформаційних технологій у своїй професійній діяльності, творчого підходу до розв'язування нестандартних задач, глибокого опанування фундаментальних основ дисциплін. Для цього було розроблено методику використання СКМ у процесі навчання дослідження операцій, спрямовану на формування системи професійних компетентностей майбутніх фахівців з інформатики, що дасть змогу у майбутньому адаптуватися до вимог інформаційного суспільства; розвиток творчого підходу до розв'язування нестандартних задач; формування математичних умінь та навичок необхідних для аналізу, моделювання та розв'язання теоретичних та прикладних задач із застосуванням СКМ [12]. Використання цієї методики постало предметом експериментального дослідження, у ході якого застосовувалася як локальна, так і хмаро орієнтована реалізація СКМ Maxima.

Однією з важливих сфер використання СКМ Maxima у наукових дослідженнях і у вивченні математичних та інформатичних дисциплін у вищій школі є розв'язування і дослідження оптимізаційних задач, що виникають у різних галузях людської діяльності тощо.

За умови запровадження СКМ Maxima у процес навчання дослідження операцій виникає можливість зосередити увагу студентів на засадничих поняттях, принципах, підходах за рахунок вивільнення часу і зусиль, які йдуть на встановлення, підтримування, обслуговування програмного забезпечення, та навіть значною мірою знівелювати реальні просторові та часові межі реалізації доступу до необхідних електронних ресурсів. Даний підхід розвиває міжпредметні зв'язки, сприяє поглибленому вивченню матеріалу, розширює можливості самостійного дослідження, поєднання теорії і практики, інтеграції знань стосовно різних підрозділів та рівнів інформатичної освіти [12; 14].

Для цього може бути застосована технологія *«віртуального робочого столу»*, за якої опрацювання і зберігання даних відбувається у центрі опрацювання даних (ЦОД), а для користувача робота з хмарними додатками, звернення до яких відбувається через Інтернет-браузер, нічим не відрізняється від роботи з програмним забезпеченням, встановленим на робочому столі його персонального комп'ютера [12; 18].

Використання програмного забезпечення, що встановлено на віртуальному робочому столі студента, не потребує витрачання навчального часу на інсталяцію і оновлення, створюються умови для більш диференційованого підходу до організації навчання, дає можливість зосередитися на вивченні основного матеріалу [12].

Необхідність використання СКМ у навчальному процесі, обумовлена ще й тим, що робота з ними надає реальну можливість студентам набути вмінь розв'язувати практичні задачі з використанням СКМ за відомою схемою: *постановка задачі → визначення цілей моделювання → побудова математичної моделі → обрання математичного методу і алгоритму розв'язування задачі → реалізація математичної моделі з використанням СКМ → проведення розрахунків → аналіз одержаних результатів та їх інтерпретація → прийняття рішення.*

У рамках дисципліни «Дослідження операцій» вивчається велика кількість практичних задач, які зручно інтерпретувати як задачі оптимізації на графах. Прикладами таких задач є відшукання найкоротшого маршруту між двома населеними пунктами, визначення максимальних пропускних характеристик нафтопроводу, укладання календарного плану виконання робіт проекту тощо.

Під час розв'язування оптимізаційних задач на графах реалізовується міжпредметні зв'язки інформатичних, математичних, економічних та інших дисциплін, що сприяє інтелектуальному розвитку студентів на основі формування уявлень про цілісність бачення світу, забезпечується формування навичок володіння не тільки декларативними, але й процедурними знаннями. Використання теорії графів до розв'язування задач формує у студентів вміння подавати умови задачі мовою теорії графів, а потім інтерпретувати отриманий розв'язок в термінах початкової задачі.

Можливості використання системи Maxima для розв'язування задач оптимізації на графах досить широкі. Студент, використовуючи СКМ Maxima, розв'язує поставлену перед ним задачу, й, отже, у нього не виникає психологічного бар'єру у застосуванні математичного апарату, а попри цн він також усвідомлює, який матеріал треба повторити (або вивчити). Розв'язування задач прикладного характеру (такими, зокрема є оптимізаційні задачі на графах) з використанням СКМ надає можливість формування професійних компетентностей. Цікавими також є дослідження задач теорії оптимізації, зокрема реалізації чисельних методів як умовної, так і безумовної оптимізації з використанням СКМ Maxima.

У вивченні розділу «Моделі динамічного програмування» студентам пропонуються для розв'язання задачі, для розв'язання яких використовуються команди та функції Maxima або створюються власні процедури та функції. Це, у свою чергу, сприяє вдосконаленню навичок програмування. Наприклад, у розв'язанні задачі динамічного програмування про рюкзак студенти виконують дослідницьку, творчу роботу, а її рутинна частина виконується за допомогою комп'ютера.

Головними етапами в розв'язуванні таких задач є постановка задачі (задання цільової функції, критерію оптимальності, обмежень, задання точності розв'язку) і дослідження отриманих результатів. У студентів формуються основи системного підходу до розв'язування задач, а також вони бачать взаємозв'язки змісту навчання різних навчальних дисциплін.

Підсумовуючи розгляд вивчення курсу «Дослідження операцій», слід зазначити, що широкий набір засобів для комп'ютерного підтримування аналітичних, обчислювальних та графічних операцій роблять системи комп'ютерної математики одними з основних засобів у професійній діяльності математиків та програмістів. Дослідження з використанням системи Maxima поєднують алгебраїчні методи з обчислювальними. У цьому розумінні СКМ – поєднуюча ланка між математикою та інформатикою, де дослідження зосереджується як на розробці алгоритмів для символьних обчислень й опрацювання даних за допомогою комп'ютера, так і на створенні програм для реалізації подібних алгоритмів.

## 3.2. Характеристики системи MAXIMA й особливості її реалізації як у локальному, так і у хмаро орієнтованому варіанті

Система Maxima працює на всіх сучасних варіантах операційних Linux та UNIX, Windows 9x/2000/XP» [14, с. 140]. Зокрема у Дрогобицькому державному педагогічному університеті імені Івана Франка була реалізована хмарна версія системи Maxima, встановлена на віртуальному сервері з операційною системою Ubuntu 10.04 (Lucid Lynks). У репозитарії цієї операційної системи є версія системи Maxima на основі редактора Emacs, що і була встановлена на віртуальний робочий стіл студента [12; 14].

Поряд з цим, всі основні дії з цим програмним продуктом можна здійснювати й у середовищі з графічним інтерфейсом wxMaxima, який базується на wxWidgets, під управлінням операційної системи Windows [12; 14].

Система Maxima серед математичних пакетів володіє досить широкими можливостями під час виконання символьних обчислень. Це, по суті, єдина з вільно поширюваних відкритих систем, яка не поступається комерційним СКМ Mathematica та Maple. Система Maxima розповсюджується під ліцензією GPL і є доступною як користувачам операційних систем Linux, так і користувачам Windows [14, с. 139].

*Робота із системою Maxima під управлінням ОС Windows.*

Система Maxima працює на всіх сучасних варіантах операційних Linux та UNIX, Windows 9x/2000/XP. Розглянемо роботу з системою Maxima з графічним інтерфейсом wxMaxima, який базується на wxWidgets, під управлінням операційної системи Windows [14, с. 140].

*Робота із системою Maxima під управлінням ОС LINUX.*

Робота із системою Maxima в ОС Linux може відбуватися в різний спосіб. Використовуючи віддалений робочий стіл на базі ОС Ulteo (дистрибутив Linux, створений на основі ОС Ubuntu) зручно використовувати середовище texmacs, який встановлюється як статичний додаток на ОС [12; 14]. ОС Ulteo може бути встановлене як на один з комп'ютерів локальної мережі, на віртуальній машині, створеній на продуктах фірми VMWare, на віртуальній машині в системі AWS або якихось інших подібних мережних системах. Тут і далі наведено досвід використання системи Maxima, що була встановлена на хмарному сервері, який більш докладно висвітлено в [12].

Texmacs передбачає роботу з декількома системами, однією з яких є система Maxima. Для створення сесії (тобто вставити об'єкт) Maxima вибираємо послідовно опції меню Insert – Session – Maxima. З'являється активний рядок для введення команд системи Maxima. Більш докладно досвід використання СКМ Maxima у хмаро орієнтованому середовищі охарактеризовано у роботі «Методичні рекомендації з використання хмаро орієнтованого компонента на базі системи Maxima у навчанні інформатичних дисциплін» [12].

Наведемо приклад лабораторного заняття з теми «Побудова каркасу мінімальної вартості. Алгоритм Прима» з використанням методів доцільно дібраних задач та демонстраційних прикладів, що розв'язується з використанням хмаро орієнтованого компонента навчального середовища на базі системи Maxima. Виконання цього завдання може бути здійснено також і з застосуванням локальної версії, тому хід розв'язання супроводжуються ілюстраціями, що можуть бути отримані у будь-якій з версій системи, а також інтерфейсами, що отримані між час реалізації хмарного компонента. Після викладу основних теоретичних положень з теми й алгоритму Прима студентам пропонується такий приклад.

**Приклад**. Побудувати каркас мінімальної вартості на основі алгоритму Прима для заданого графа. На рис. 1 зображено заданий граф і відповідну матрицю ваг.

Матриця ваг матиме вигляд:

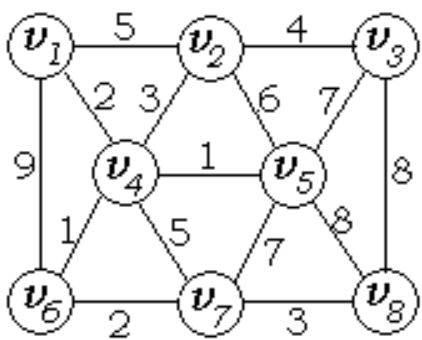

| | 1 | 2 | 3 | 4 | 5 | 6 | | 8 |
|---|---|---|---|---|---|---|---|---|
| **1** | 0 | 5 | * | 2 | * | 9 | * | * |
| **2** | 5 | 0 | 4 | 3 | 6 | * | * | * |
| **3** | * | 4 | 0 | * | 7 | * | * | 8 |
| **4** | 2 | 3 | * | 0 | 1 | 1 | 5 | * |
| **5** | * | 6 | 7 | 1 | 0 | * | 7 | 8 |
| **6** | 9 | * | * | 1 | * | 0 | 2 | * |
| **7** | * | * | * | 5 | 7 | 2 | 0 | 3 |
| **8** | * | * | 8 | * | 8 | * | 3 | 0 |

*Рис. 1. Задання графа умови задачі та побудова матриці ваг*

Далі студентам пропонується демонстраційний приклад (програмна реалізація даного завдання у середовищі СКМ Maxima).

**ПРОГРАМА 1**

```
c: genmatrix(lambda([i,j], inf), n, n)$
for i:1 thru n do
   (for j:1 thru n do
       (if member([i,j],edges(g)) then (c[i,j]:get_edge_weight([i,j],g),
                                         c[j,i]:get_edge_weight([i,j],g))
)

   )

   )$
   c;
   V:vertices(g)$
   w:1$
   W:delete(w,V)$
   T:[]$
   for v:1 thru n do
     (near[v]:w,
      d[v]:c[v,w])$
   for i:1 while i<n do
       (
   dmin:inf,
         for j:1 thru n do
         (if (d[j]<dmin) and (member(j,W)) then (v:j,dmin:d[j])),
   T:append(T,[[near[v],v]]),

W:delete(v,W),
for u:1 thru n do

(if (d[u]>c[u,v]) and (member(u,W))
          then (near[u]:v,d[u]:c[u,v]))
)$
```

$$\begin{bmatrix} \infty & 5 & \infty & 2 & \infty & 9 & \infty & \infty \\ 5 & \infty & 4 & 3 & 6 & \infty & \infty & \infty \\ \infty & 4 & \infty & \infty & 7 & \infty & \infty & 8 \\ 2 & 3 & \infty & \infty & 1 & 1 & 5 & \infty \\ \infty & 6 & 7 & 1 & \infty & \infty & 7 & 8 \\ 9 & \infty & \infty & 1 & \infty & \infty & 2 & \infty \\ \infty & \infty & \infty & 5 & 7 & 2 & \infty & 3 \\ \infty & \infty & 8 & \infty & 8 & \infty & 3 & \infty \end{bmatrix}$$

```
GG:create_graph([1,2,3,4,5,6,7,8],T)$
draw_graph(GG,show_id=true)$
```

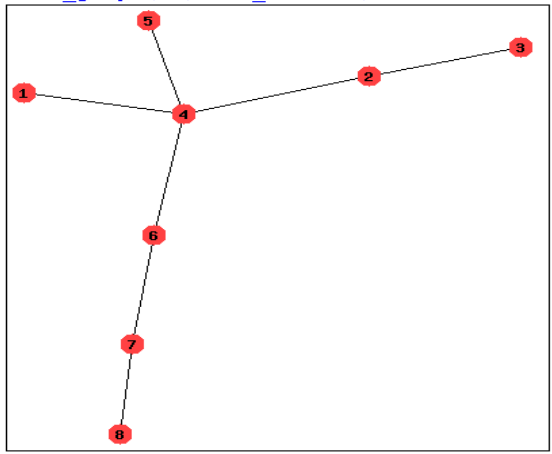

*Рис. 2. Результат побудови графа*

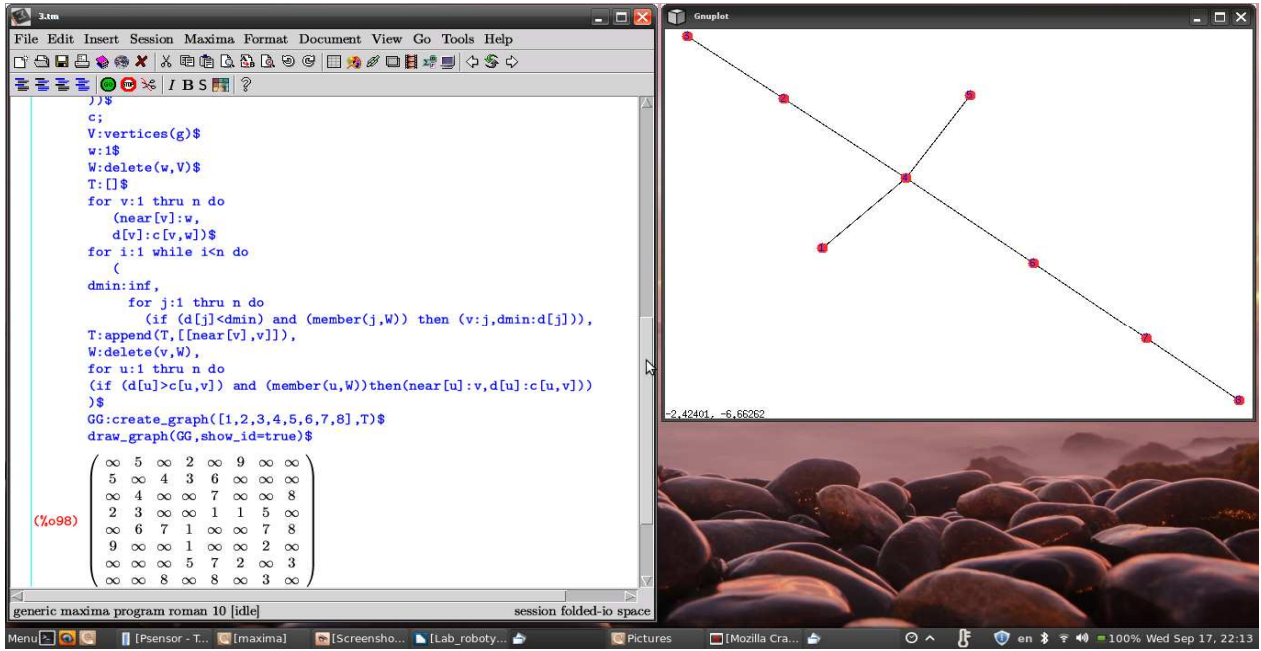

*Рис. 3. Результат побудови графа за допомогою хмаро орієнтованого компонента*

Н а рис.2 зображено результат побудови графа, а на рис. 3 – вигляд інтерфейсу під час розв'язування цього завдання з використанням хмаро орієнтованого компонента.

Після цього пропонуємо перевірити результат, використавши функцією побудови каркасу мінімальної вартості minimum_spanning_tree(g).

```
load(graphs)$

g:create_graph([1,2,3,4,5,6,7,8],
                 [[[1,2],5], [[1,4],2], [[1,6],9],
                   [[2,3],4],[[2,4],3],[[2,5],6],
                   [[3,5],7],[[3,8],8],
                     [[4,5],1],[[4,6],1],[[4,7],5],
                       [[5,7],7],[[5,8],8],
                         [[6,7],2],
                           [[7,8],3]
                 ])$
draw_graph(g,show_weight=true,show_id=true)$
```

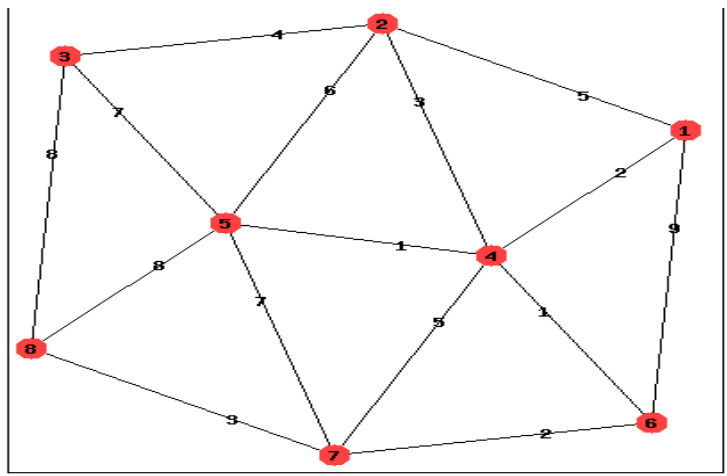

*Рис. 4. Результат побудови графа*

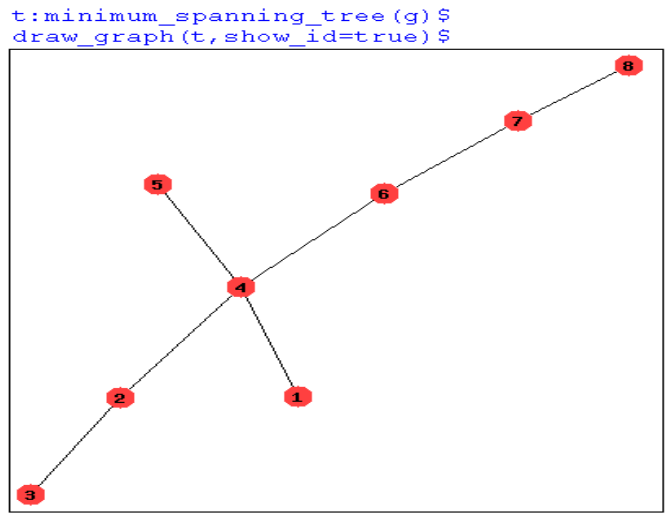

*Рис. 5. Побудова каркаса мінімальної вартості*

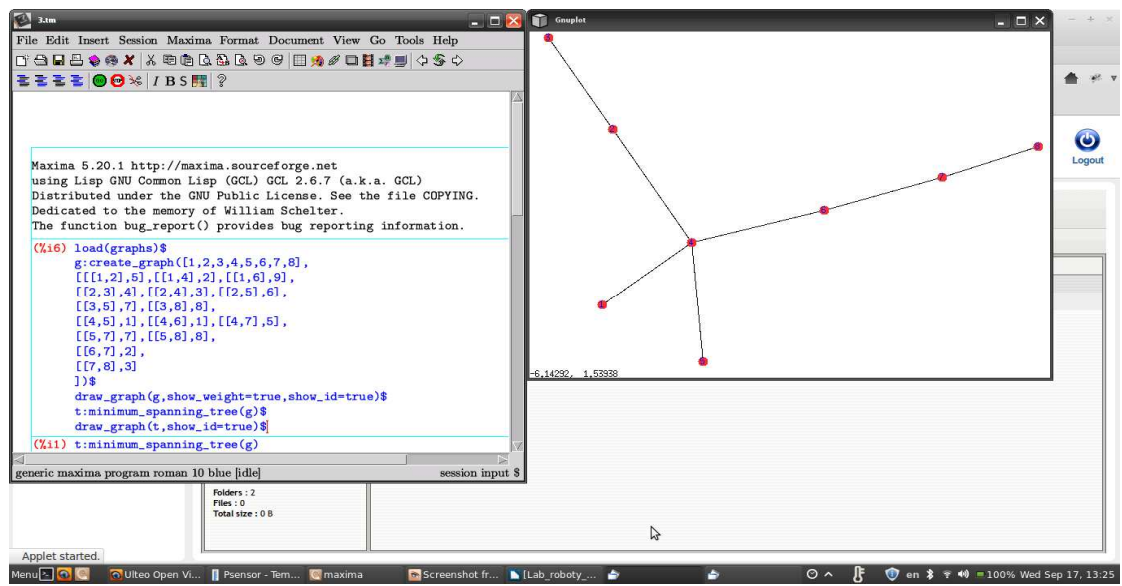

*Рис. 6. Побудова каркаса мінімальної вартості за допомогою хмаро орієнтованого компонента*

На рис. 4 показано результат побудови графа, на рис. 5 – результат побудови каркаса мінімальної ваги, на рис. 6 – результат цієї ж дії з використанням хмаро орієнтованого компонента.

У кінці кожної лабораторної роботи пропонуються завдання для самостійної роботи.

З наведеного прикладу видно, що завдяки застосуванню СКМ Maxima у процесі навчання дослідження операцій виникає можливість здійснити необхідні обчислення, надаючи можливість студентам більшу частину навчального часу використати для дослідження методів розв'язування прикладних задач чи навіть їх розробки, набуття навичок побудови математичних моделей, інтерпретації та аналізу результатів обчислювального експерименту, що призводить до більш глибокого розуміння фундаментальних понять, які вивчаються. Використання СКМ Maxima надає можливість забезпечити повноцінну навчально-пізнавальну, методичну і науково-дослідну діяльність, вводити інновації в навчальний процес, реалізовувати принцип міжпредметності, поєднувати самостійну роботу з різними формами колективної діяльності.

### 3.3. Результати педагогічного експерименту з використання системи Maxima у процесі навчання дослідження операцій

Протягом 2010–2014 р.р. проводилося експериментальне дослідження, у ході якого здійснювалося запровадження СКМ MAXIMA у процес навчання дослідження операцій студентів Інституту фізики, математики та інформатики Дрогобицького державного педагогічного університету імені Івана Франка (ДДПУ) ОКР «Бакалавр» галузі знань 0403 «Системні науки та кібернетика» напряму підготовки 6.040302 «Інформатика". В експерименті з апробування спеціально розробленої методики навчання дослідження операцій з використанням системи Максима. В експерименті, на його формувальному етапі, взяли участь 240 студентів. Експеримент підтвердив гіпотезу дослідження, щодо підвищення рівня сформованості фахових компетентностей у процесі навчання згідно розробленої методики, а також показав, що завдяки засобам хмарних технологій можна досягти розширення доступу до засобів дослідницької діяльності студентів [13].

В експерименті були задіяні як локальна версія системи, встановлена на робочому столі у комп'ютері студента, так і хмарна версія, що була розміщена на віртуальному робочому столі.

У ході експерименту були визначені показники експертного оцінювання доцільності запровадження у навчальний процес і використання розробленого хмаро орієнтованого компонента. Для цього було використано метод оцінювання якості електронних освітніх ресурсів, що був розроблений у межах діяльності спільної науково-дослідної лабораторії Інституту інформаційних технологій і засобів навчання НАПН України та Херсонського державного університету «Управління якістю засобів ІКТ навчального призначення» [22], адаптований і застосований для випадку експертного оцінювання хмаро орієнтованого засобу. Було спеціально відібрано 20 експертів, що мали досвід навчання математичних та інформатичних дисциплін у вищому навчальному закладі з використанням ІКТ, і були досить обізнаними з особливостями використання хмарних технологій, або ж застосовували їх у навчальній діяльності [5; 16].

До розробленої системи показників і відповідних параметрів їх оцінювання входило дві їх групи. Перша група охоплювала 7 техніко-технологічних показників: зручність організації доступу; зрозумілість інтерфейсу; надійність; зручність підтримування колективної роботи; легкість інтеграції з іншими ресурсами в єдине середовище; мобільність (доступність з будь-якого пристрою); потрібність (корисність) [16].

Друга група містила 9 психолого-педагогічних показників: науковість; доступність; розвиток інтелектуального потенціалу того, хто вчиться; проблемність; індивідуалізація; адаптивність; методична доцільність; професійна орієнтованість; наявність зворотного зв'язку [16].

Треба зазначити, що техніко-технологічні показники, що характеризують якість хмаро орієнтованого ЕОР, є досить специфічними, їх визначення особливо суттєво для оцінювання хмаро орієнтованого ресурсу, тому було приділено особливу увагу їх визначенню. У той же час, психолого-педагогічні показники, зазвичай, не є специфічними, вони використовуються для оцінювання будь-яких видів ЕОР.

Було здійснено валідизацію показників, у ході чого експертам було запропоновано оцінити значущість кожного показника для оцінювання хмаро орієнтованого ЕОР. В опитуванні взяли участь 20 експертів, що оцінювали валідність обраних 16 показників.

Експертне оцінювання якості ЕОР може вважатися достатньо достовірним лише у випадку гарної узгодженості відповідей експертів. Тому статистичне опрацювання результатів експертного оцінювання має охоплювати аналіз погодженості думок експертів. Метод конкордації використовується для того, щоб оцінити ступінь погодження експертів за такими чинниками: зважування типів ЕОР, параметризація показників якості ЕОР, загальна оцінка якості ЕОР [22; с. 322].

Другим етапом було ранжування, у результаті якого показникам були присвоєні рангові значення, що характеризують їхнє місце в порядку зростання їх значущості. В опитуванні взяли участь 19 експертів, що оцінювали валідність обраних 16 показників [16]. Результати визначення рангових значень параметрів валідності ЕОР наведено в таблиці 1.

*Таблиця 1*

**Рангові значення параметрів, що характеризують валідність показників якості ЕОР**

| Експерт | Рангові значення, що характеризують валідність показників якості ЕОР | | | | | | | | | | | | | | | |
|---|---|---|---|---|---|---|---|---|---|---|---|---|---|---|---|---|
| | **1** | **2** | **3** | **4** | **5** | **6** | **7** | **8** | **9** | **10** | **11** | **12** | **13** | **14** | **15** | **16** |
| 1 | 1 | 2 | 3 | 4 | 5 | 6 | 7 | 8 | 9 | 10 | 11 | 12 | 13 | 14 | 15 | 16 |
| 2 | 9 | 13 | 14 | 10 | 5 | 1 | 6 | 3 | 11 | 15 | 16 | 12 | 7 | 2 | 8 | 4 |
| 3 | 7 | 1 | 2 | 9 | 10 | 3 | 11 | 15 | 8 | 4 | 5 | 12 | 13 | 6 | 14 | 16 |
| 4 | 1 | 2 | 3 | 4 | 5 | 6 | 7 | 8 | 9 | 10 | 11 | 12 | 13 | 14 | 15 | 16 |
| 5 | 1 | 3 | 5 | 6 | 7 | 8 | 9 | 10 | 2 | 4 | 11 | 12 | 13 | 14 | 16 | 16 |
| 6 | 11 | 7 | 12 | 13 | 8 | 1 | 2 | 3 | 14 | 9 | 15 | 16 | 10 | 4 | 5 | 6 |
| 7 | 13 | 11 | 1 | 2 | 3 | 4 | 5 | 14 | 15 | 12 | 6 | 7 | 8 | 9 | 10 | 16 |
| 8 | 5 | 6 | 1 | 7 | 2 | 8 | 9 | 10 | 11 | 12 | 3 | 13 | 4 | 14 | 15 | 16 |
| 9 | 15 | 11 | 4 | 6 | 5 | 2 | 10 | 14 | 16 | 12 | 3 | 7 | 9 | 1 | 8 | 13 |
| 10 | 11 | 10 | 1 | 4 | 12 | 9 | 7 | 6 | 5 | 8 | 3 | 2 | 16 | 14 | 13 | 15 |
| 11 | 1 | 2 | 3 | 4 | 5 | 6 | 7 | 8 | 9 | 10 | 11 | 12 | 13 | 14 | 15 | 16 |
| 12 | 12 | 14 | 13 | 10 | 6 | 2 | 7 | 4 | 11 | 16 | 15 | 9 | 5 | 1 | 8 | 3 |
| 13 | 8 | 1 | 3 | 9 | 10 | 2 | 12 | 15 | 7 | 5 | 4 | 11 | 13 | 6 | 14 | 16 |
| 14 | 8 | 1 | 7 | 2 | 11 | 9 | 3 | 10 | 12 | 4 | 15 | 13 | 14 | 5 | 16 | 6 |
| 15 | 1 | 4 | 11 | 8 | 7 | 10 | 6 | 13 | 2 | 3 | 5 | 9 | 12 | 14 | 15 | 16 |
| 16 | 14 | 8 | 12 | 11 | 7 | 2 | 3 | 1 | 13 | 10 | 15 | 16 | 9 | 4 | 5 | 6 |
| 17 | 14 | 11 | 2 | 1 | 4 | 5 | 3 | 13 | 15 | 12 | 6 | 7 | 9 | 8 | 10 | 16 |
| 18 | 13 | 11 | 3 | 10 | 1 | 12 | 9 | 7 | 6 | 8 | 2 | 5 | 4 | 15 | 14 | 16 |
| 19 | 15 | 11 | 3 | 7 | 5 | 1 | 10 | 12 | 16 | 14 | 4 | 6 | 9 | 2 | 8 | 13 |

Коефіцієнт конкордації W обчислено згідно запропонованої М. Кендлом формули [21]:

$$W = \frac{12S}{m^2(n^3 - n)} \qquad (1)$$

$$\text{де} \qquad S = \sum_{i=1}^{n} \{ \sum_{j=1}^{m} x_{ij} \}^2 , \qquad (2)$$

$m$ – кількість експертів, $n$ – кількість об'єктів оцінювання (якість параметрів), $x_{ij}$ – оцінка $i$-об'єкта $j$-експертом. Коефіцієнт конкордації повинен знаходитися в межах від 0 до 1. Якщо $W = 1$, це означає, що всі експерти дають однакові оцінки всім параметрам, якщо $W = 0$, оцінки експертів не узгоджуються.

Використовуючи формулу (1), ми визначаємо, що коефіцієнт $W = 0.189$ і він помітно відрізняється від нуля, так ми можемо припустити, що серед експертів існують об'єктивні узгодження. Враховуючи, що значення

$m(n-1)W$ розподіляється відповідно до $\chi^2$ з $(n-1)$ ступенем свободи, тоді

$$\chi_W^2 = \frac{12S}{mn(n+1)} = 52,8 \qquad (3)$$

Порівнюючи це значення з табличним значенням $\chi_T^2$ для $n - 1 = 15$ ступеня свободи і рівня значущості $\alpha = 0,01$, ми знаходимо

$$\chi_W^2 = 52.8 > \chi_T^2 = 30,5 \qquad (4)$$

Отже, гіпотеза про узгодженість експертних оцінок підтверджується відповідно з Пірсоном.

Таким чином, результати опитування підтвердили припущення, що метод експертного оцінювання може бути застосований для оцінювання якості електронних освітніх ресурсів у хмаро орієнтованому середовищі [16].

Отже, проблема була поставлена так: чи доцільно і доречно спроектовано досліджувані ЕОР у хмаро орієнтованому середовищі? Або інакше кажучи: чи відповідає певним вимогам до якості запропонований хмаро орієнтований компонент навчального середовища?

Для цієї мети було розроблено 2 анкети, за якими експерти мали б оцінити згаданий компонент за 2 групами показників. Так, 20 експертів оцінили 16 параметрів (серед них 7 – техніко-технологічних і 9 – психолого-педагогічних). Чотирирівнева шкала була використана для вимірювання значень параметрів: 0 (не виявлений), 1 (низький), 2 (добрий), 3 (дуже добрий).

Результуючі середні значення оцінювання, отримані за кожним з техніко-технологічних показників, були такі: «Зручність доступу» = 2.1, «Зрозумілість інтерфейсу» = 2.4, «Швидкодія» = 2.1, «Надійність при роботі через браузер» = 2.56, «Підтримка колективної роботи» = 2.0, «Зручність інтеграції» = 2.0, «Корисність» = 2.8. Вага всіх критеріїв була прийнята за 1, і отримано результуюче значення: 2.3.

Результуючі значення оцінювання за кожним з психолого-педагогічних показників, були такі: «Науковість» = 2.6, «Доступність» = 2.7, «Розвиток інтелектуального потенціалу» = 2.5, «Проблемність» = 2.8, «Індивідуалізація» = 2.8, «Адаптивність» = 2.6, «Методична доцільність» = 2.81. «Професійна орієнтованість» = 2,75, «Зворотній зв'язок» = 2,75. Загальний показник: 2.71.

Більш докладно результати валідації показників експертного оцінювання розроблено хмаро орієнтованого компонента, методику оцінювання і результати роботи експертів висвітлено в [23].

Отримані результати свідчать, що даний програмний продукт має достатньо високі значення показників оцінювання, що свідчить про доцільність його подальшого впровадження і використання в навчальному процесі [16]. Про покращення щодо реалізації дослідницького підходу до навчання, зокрема, завдяки проведенню комп'ютерних експериментів у середовищі СКМ, свідчать досить високі показники за низкою психолого-педагогічних показників, зокрема, «Проблемність» = 2.8, «Індивідуалізація» = 2.8, «Адаптивність» = 2.6, «Методична доцільність» = 2.81.

«Професійна орієнтованість» = 2,75, «Зворотній зв'язок» = 2,75. Загальний показник: 2.71.

Перевагою даного підходу до оцінювання якості є можливість порівнювати різні шляхи організації доступу до електронних ресурсів у хмаро орієнтованому середовищі, визначати кращі рішення щодо проектування середовища й організації його сервісів. За даними параметрами можна проводити експертне оцінювання різних типів хмаро орієнтованих компонентів, розроблених як на базі гібридної або корпоративної хмари, так і програмного забезпечення, що постачається за моделлю «програмне забезпечення як сервіс».

У ході впровадження хмаро орієнтованого компоненту проводилося також навчання викладачів через систему тренінгів. В експериментальній групі було 32 учасника, у контрольній групі – 60 учасників. Частка учасників з високим рівнем ІКТ компетентності зросла з 21% до 53% в експериментальній групі; з 20% до 33% у контрольній групі. Значення критерію Фішера $\varphi_{емп} = 1,83 \succ \varphi_{0,05} = 1,64$, це означає, що відмінність є статистично значущою [16].

## 4. ВИСНОВКИ ТА ПЕРСПЕКТИВИ ПОДАЛЬШИХ ДОСЛІДЖЕНЬ

Результати педагогічного експерименту підтвердили, що досліджувані хмаро орієнтовані компоненти характеризують достатньо високі значення психолого-педагогічних і техніко-технологічних показників.

Результати експериментального дослідження підтверджують гіпотезу про те, що методично обґрунтоване і педагогічно виважене використання хмарних сервісів підтримування науково-педагогічної діяльності у вищому навчальному закладі на основі спеціально розробленої методики сприятиме зростанню рівня ІКТ-компетентності студентів і викладачів, розширенню доступу до електронних ресурсів і сервісів, покращенню організації навчання.

СКМ є середовищем для проектування засобів навчання, тому можуть бути використані для створення інноваційних педагогічних технологій. Перспективним напрямом є використання засобів даного типу «у хмарі», коли виникає більше можливостей адаптації середовища навчання до навчальних потреб користувача.

Упровадження СКМ у процес навчання майбутніх фахівців з інформатики надає можливість активізувати навчально-пізнавальну активність студентів, сприяє розвитку їхніх творчих здібностей, математичної інтуїції та навичок здійснення дослідницької діяльності. Систематичне використання СКМ, сприяє формуванню у студентів ставлення до комп'ютера як до засобу розв'язування професійних задач. Такі студенти отримують більш глибокі знання не тільки з математичних дисциплін, але й з інформатики. Як правило, у них немає психологічного бар'єру перед використанням складних програмних засобів. Навпаки, їх приваблюють створені на високому професійному рівні програми, і вони помічають унікальні можливості застосування таких систем.

Перспективою подальших досліджень є розширення кола дослідницьких задач, що можна розв'язувати з використанням запропонованого хмаро орієнтованого компонента, його подальше випробування, а також порівняння з іншими хмаро орієнтованими програмними продуктами на основі визначеної системи показників.

## СПИСОК ВИКОРИСТАНИХ ДЖЕРЕЛ

1. Бабій Ю. О. Хмарні обчислення проти розподілених обчислень: сучасні перспективи / Ю. О. Бабій, В. П. Нездоровін, Є. Г. Махрова, Л. П. Луцкова // Вісник Хмельницького національного університету. Технічні науки. — 2011. — № 6. — С. 80–85.
2. Биков В. Ю. Методологічні та методичні основи створення і використовування електронних засобів навчального призначення / В. Ю. Биков, В. В. Лапінський // Комп'ютер у школі та сім'ї. — 2012. — №2(98). — С. 3–6.
3. Биков В. Ю. Відкрита освіта і відкрите навчальне середовище / В. Ю. Биков // Теорія і практика управління соціальними системами. — 2008. — №2. — С. 116–123.
4. Биков В. Ю. Технології хмарних обчислень, ІКТ-аутсорсінг та нові функції ІКТ-підрозділів навчальних закладів і наукових установ / В. Ю. Биков // Інформаційні технології в освіті. — Вип. 10. — Херсон : ХДУ, 2011. — № 10. — С. 8–23.
5. Бугаєць Н.О. Моделювання анімаційних наочностей засобами графічного середовища програми Maxima / Бугаєць Н.О. Інформаційні технології і засоби навчання. — 2015. — 3 (47). — С. 67–79.
6. Дем'яненко В. М. Дослідно-експериментальна діяльність Інституту інформаційних технологій і засобів навчання НАПН України на базі навчальних закладів різних рівнів / В. М. Дем'яненко, Ю. Г. Носенко, О. П. Пінчук, М. П. Шишкіна // Комп'ютер у школі та сім'ї. — 2015. — № 5 (125). — С. 18–23.
7. Єрохін С. Технологічні уклади, динаміка цивілізаційних структур та економічна перспектива України / С. Єрохін // Економічний часопис-ХХІ. — 2006. — № 1–2.
8. Кобильник Т. П. Системи комп'ютерної математики: Maple, Mathematica, Maxima / Т. П. Кобильник. — Дрогобич : Редакційно-видавничий відділ ДДПУ імені Івана Франка, 2008. — 316 с.
9. Словак К. І. Мобільні математичні середовища: сучасний стан та перспективи розвитку / К. І. Словак, С. О. Семеріков, Ю. В. Триус // Науковий часопис НПУ імені М. П. Драгоманова. Серія 2. Комп'ютерно-орієнтовані системи навчання : зб. наук. пр. — К. : НПУ ім. М. П. Драгоманова. 2012. — № 12 (19). — С. 102–109.
10. Триус Ю. В. Комп'ютерно-орієнтовані методичні системи навчання математичних дисциплін у вищих навчальних закладах : дис. ... докт. пед. наук : спец. 13.00.02 / Юрій Васильович Триус. — Черкаси, 2005. — 649 с.
11. Триус Ю. В. Використання WEB-СКМ у навчанні методів оптимізації та дослідження операцій студентів математичних та комп'ютерних спеціальностей/ Ю. В.Триус // Інноваційні комп'ютерні технології у вищій школі : матеріали 4-ої наук.-прокт. конференції / Національний університет «Львівська політехніка». — Львів : В-во Львівська політехніка, 2012. — С. 110–115
12. Шишкіна М. П. Методичні рекомендації з використання хмаро орієнтованого компонента на базі системи Maxima у навчанні інформатичних дисциплін / М. П. Шишкіна, У. П. Когут. – Дрогобич : Ред.-вид. відділ ДДПУ ім. І. Франка, 2014. – 57 с.
13. Шишкіна М. П. Перспективи розвитку освітнього середовища та підвищення якості інноваційних засобів ІКТ / М. П. Шишкіна // Гуманітарний вісник ДВНЗ «Переяслав-Хмельницький державний педагогічний університет імені Григорія Сковороди» — Тематичний випуск «Вища освіта України у контексті інтеграції до європейського освітнього простору». — К. : Гнозис, 2013. — Дод. 1 до Вип. 31, Том IV (46). — С. 440–446.
14. Шишкіна М. П. Формування фахових компетентностей бакалаврів інформатики у хмаро орієнтованому середовищі педагогічного університету / М. П. Шишкіна, У. П. Когут, І. А. Безвербний // Проблеми підготовки сучасного вчителя. — Умань : ФОП Жовтий О.О., 2014. — Вип. 9. — Ч. 2. — С. 136–146.
15. Шишкіна М. П. Хмаро орієнтоване середовище навчального закладу: сучасний стан і перспективи розвитку досліджень [Електрон. ресурс] / М. П. Шишкіна, М. В. Попель // Інформаційні технології і засоби навчання. — 2013. — 5 (37). — Режим доступу : http://journal.iitta.gov.ua/index.php/itlt/article/view/903/676.
16. Шишкіна М. П. Формування і розвиток хмаро орієнтованого освітньо-наукового середовища вищого навчального закладу : монографія / М. П. Шишкіна. — – К. : УкрІНТЕІ, 2015. — 256 с.
17. Шишкіна М.П. Тенденції розвитку і стандартизації вимог до засобів ІКТ навчального призначення на базі хмарних обчислень / М.П.Шишкіна // Науковий вісник Мелітопольського державного педагогічного університету. Серія: Педагогіка. — Вип.2(13). — 2014. — С. 223–231.

18. Шишкіна М. П. Формування і розвиток засобів ІКТ освітньо-наукового середовища вищого навчального закладу на базі концепції хмарних обчислень / М. П. Шишкіна // Гуманітарний вісник ДВНЗ «Переяслав-Хмельницький державний педагогічний університет імені Григорія Сковороди». Тематичний випуск «Вища освіта України в контексті інтеграції до європейського освітнього простору». – К. : Гнозис, 2014. – Дод. 1 до Вип. 5, Т. III (54). – С. 302–309.

19. Шишкіна М. П. Моделі організації доступу до програмного забезпечення у хмаро орієнтованому освітньому середовищі / М. П. Шишкіна // Інформаційні технології в освіті. – № 22. – 2015. – С. 120–129.

20. Cusumano M. Cloud computing and SaaS as new computing platforms / Michael Cusumano // Communications of the ACM. – 53(4). – 2010. – Pp. 27–29.

21. Kendall M. Rank Correlation Methods / M. Kendall. – London: Charles Griffen & Company, 1948.

22. Kravtsov H. M. Methods and Technologies for the Quality Monitoring of Electronic Educational Resources [Electronic resource] / H. M. Kravtsov // S. Batsakis et al. (eds.) Proc. 11-th Int. Conf. ICTERI 2015, Lviv, Ukraine, May 14-16. – 2015. – P. 311–325. – CEUR-WS.org/ Vol. 1356, ISSN 1613–0073, P. 311–325. – Available at: CEUR–WS.org/Vol–1356/paper_109.pdf

23. Mariya Shyshkina. The Hybrid Cloud-based Service Model of Learning Resources Access and its Evaluation / [Електронний ресурс] // Proceedings of the 12th International Conference on ICT in Education, Research and Industrial Applications. Integration, Harmonization and Knowledge Transfer / Ed. by Sotiris Batsakis, Heinrich C. Mayr, Vitaliy Yakovyna. – CEUR Workshop Proceedings. - vol.1356. – 2015. – Pp. 295–310. – Режим доступу : http://ceur-ws.org/Vol-1614/paper_57.pdf.

24. Mell P. The NIST Definition of Cloud Computing. Recommendations of the National Institute of Standards and Technology / P.Mell, T.Grance. – NIST Special Publication 800-145. NIST, Gaithersburg, MD 20899-8930, September 2011.

25. Shyshkina M. Emerging Technologies for Training of ICT-Skilled Educational Personnel / M. Shyshkina // Information and Communication Technologies in Education, Research, and Industrial Applications / V. Ermolayev, H. C. Mayr, M. Nikitchenko, A. Spivakovsky, G. Zholtkevych (Eds.). – Springer International Publishing, 2013. – P. 274–284.

26. Shyshkina M.P. Prospects of the Development of the Modern Educational Institutions' Learning and Research Environment: to the 15th Anniversary of the Institute of Information Technologies and Learning Tools of NAPS of Ukraine / M.P. Shyshkina, Y. G. Zaporozhchenko, H. M. Kravtsov // Information technologies in education. – Kherson, 2014. – № 19. – P. 62–70.

27. Turner M. Turning software into a service / M. Turner, D. Budgen, P. Brereton // Computer. – 36 (10). – 2003. – Pp. 38-44.

28. Vaquero L. M. EduCloud: PaaS versus IaaS cloud usage for an advanced computer science course / Vaquero Luis M. // IEEE Transactions on Education. – 54(4). – 2011. – Pp. 590–598.

29. Wick D. Free and open-source software applications for mathematics and education / D. Wick // Proceedings of the twenty-first annual international conference on technology in collegiate mathematics. – 2009. – Pp. 300–304.

# ИСПОЛЬЗОВАНИЕ ОБЛАКО ОРИЕНТИРОВАННОГО КОМПОНЕНТА НА БАЗЕ СИСТЕМЫ MAXIMA В ПРОЦЕССЕ ОБУЧЕНИЯ ИССЛЕДОВАНИЯ ОПЕРАЦИЙ

**Шишкина Мария Павловна**
доктор педагогических наук, старший научный сотрудник
Институт информационных технологий и средств обучения НАПН Украины, г. Киев, Украина
*shyshkina@iitlt.gov.ua*

**Когут Ульяна Петровна**
кандидат педагогических наук, доцент кафедры информатики и вычислительной математики
Дрогобычский государственный педагогический университет имени Ивана Франко,
г. Дрогобыч, Украина
*ulyana_kogut@mail.ru*

**Аннотация.** В статье исследованы проблемы использования систем компьютерной математики (СКМ) в современной высокотехнологической среде, освещены перспективные пути внедрения облачно ориентированных компонентов на базе СКМ, что является существенным фактором расширения доступа к ним как к средствам учебной и исследовательской деятельности в процессе обучения исследования операций. Определена роль СКМ в подготовке бакалавров по информатике и особенности их педагогического применения в обучении исследования операций. Рассмотрены основные характеристики СКМ MAXIMA и пути организации доступа к ней как в локальной, так и в облако ориентированной реализации. Приведены результаты экспертной оценки облака ориентированного компонента на базе системы MAXIMA в учебный процесс исследования операций.

**Ключевые слова**: облачные технологии; образовательная среда; ВУЗ; гибридная сервисная модель; электронные образовательные ресурсы.

# THE USE OF THE CLOUD-BASED LEARNING COMPONENT WITH THE MAXIMA SYSTEM FOR TEACHING OPERATIONS RESEARCH

**Mariya P. Shyshkina**
Doctor of Pedagogical Sciencies, Senior Researcher
Institute of Information Technologies and Learning Tools of NAES of Ukraine, Kyiv, Ukraine
*shyshkina@iitlt.gov.ua*

**Ulyana P. Kohut**
PhD (Pedagogical Sciences), senior lecturer of the
Department of computer science and computational mathematics
Ivan Franko Drohobych State Pedagogical University, Drohobych, Ukraine
*ulyana_kogut@mail.ru*

**Abstract.** The article highlights the problem of the use of computer mathematics (SCM) in today's high-tech environment, in particular, it describes the promising ways of introducing the cloud-oriented components based on SCM, which is a significant factor in increasing access to them as a means of educational and research activities in the field of mathematics and computer science disciplines. The role of SCM in training of bachelors of computer science and especially their use in operations research learning is revealed. The main characteristics of SCM MAXIMA and ways of access to it both locally and in the cloud-oriented implementation are considered. The results of the expert evaluation of the cloud-based learning component with the use of MAXIMA system for operations research learning are presented.

**Keywords**: cloud computing technologies; learning environment; higher education institution; the hybrid service model; electronic educational resources.

## REFERENCES (TRANSLATED AND TRANSLITERATED)

1. Babiy Yu. O. Cloud computing vs distributed computing: current perspectives / Yu. O. Babiy, V. P. Nezdorovin, Ye. H. Makhrova, L. P. Lutskova // Visnyk Khmelnytskoho natsionalnoho universytetu. Tekhnichni nauky. – 2011. – № 6. – P. 80–85. (in Ukrainian)
2. Bykov V.Iu. Methodological and methodical bases of creation and use of electronic tools for educational purposes / V.Iu.Bykov, V.V. Lapinskyi // Kompiuter u shkoli ta simi №2(98). – 2012. – P.3-6. (in Ukrainian)
3. Bykov V.Yu. Open education and open learning environment / V.Iu.Bykov // Theory and practice of social systems management. – 2008. – №2. – C. 116-123. (in Ukrainian)
4. Bykov V. Yu. Cloud computing technologies, ICT outsourcing and new features of ICT departments of educational institutions and research centers / V. Yu. Bykov // Informatsiini tekhnolohii v osviti. – 2011. – № 10. – P. 8–23. (in Ukrainian)
5. Buhaiets N.O. Modeling of animation visualization by means of graphical environment tools of the program Maxima / Buhaiets N.O. Informatsiini tekhnolohii i zasoby navchannia. – 2015. – 3 (47). – P.67-79. (in Ukrainian)

6. Demianenko V. M. Experimental work of the Institute of Information Technologies and Learning Tools of NAES of Ukraine at the educational institutions of different levels / V. M. Demianenko, Yu. H. Nosenko, O. P. Pinchuk, M. P. Shyshkina // Kompiuter u shkoli ta simi. – 2015. – № 5 (125). – P. 18–23. (in Ukrainian)

7. Ierokhin S. Technological modes, civilizations structures dynamics and economic perspectives of Ukraine / S. Yerokhin // Ekonomichnyi chasopys-KhKhI. – 2006. – № 1-2. (in Ukrainian)

8. Kobylnyk T. P. Systems of computer mathematics: Maple, Mathematica, Maxima / T. P. Kobylnyk. – Drohobych : Redaktsiino-vydavnychyi viddil DDPU imeni Ivana Franka, 2008. – 316 p. (in Ukrainian)

9. Slovak K. I. Mobile mathematical environment: current state and development prospects / K. I. Slovak, S. O. Semerikov, Yu. V. Tryus // Naukovyi chasopys NPU imeni M. P. Drahomanova. Seriia 2. Kompiuterno-oriientovani systemy navchannia : zb. nauk. pr. – K. : NPU im. M. P. Drahomanova, 2012. – № 12 (19). – P. 102–109. (in Ukrainian)

10. Tryus Yu. V. Computer-oriented methodological training system of mathematical disciplines in higher education: dys. ... dokt. ped. nauk : spets. 13.00.02 / Yurii Vasylovych Tryus. – Cherkasy, 2005. – 649 p. (in Ukrainian)

11. Tryus Yu. V. Using WEB-SCM in optimization and operations research methods training of students of mathematics and computer science disciplines / Yu. V.Tryus // Innovative computer technologies in higher education : materials of the $4^{th}$ scientific and pracyical conference / Natsionalnyi universytet «Lvivska politekhnika», – Lviv : V-vo Lvivska politekhnika, 2012, – P. 110-115. (in Ukrainian)

12. Shyshkina M. P. Guidelines on the use of the cloud-based component with the Maxima system in the process of computer science courses learning / M. P. Shyshkina, U. P. Kohut. – Drohobych : Red.-vyd. viddil DDPU im. I. Franka, 2014. – 57 p. (in Ukrainian)

13. Shyshkina M. P. Prospects of educational environment development and quality increase of innovative ICT / M. P. Shyshkina // Humanitarnyi visnyk DVNZ «Pereiaslav-Khmelnytskyi derzhavnyi pedahohichnyi universytet imeni Hryhoriia Skovorody» – Tematychnyi vypusk «Vyshcha osvita Ukrainy u konteksti intehratsii do yevropeiskoho osvitnoho prostoru». – K. : Hnozys, 2013. – n. 31, vol. IV (46). – App. 1. – P. 440–446. (in Ukrainian) computer science

14. Shyshkina M. P. Formation of professional competence of computer science bachelors in the cloud-based environment of pedagogical university / M. P. Shyshkina, U. P. Kohut, I. A. Bezverbnyi // Problemy pidhotovky suchasnoho vchytelia. – Uman: FOP Zhovtyi O.O., 2014. – n. 9, Ch. 2. – P. 136–146. (in Ukrainian)

15. Shyshkina M. P. Cloud-based environment of educational institution: current state and prospects of research [online] / M. P. Shyshkina, M. V. Popel // Informatsiini tekhnolohii i zasoby navchannia. – 2013. – 5 (37). – Available from:http://journal.iitta.gov.ua/index.php/itlt/article/view/903/676 (in Ukrainian)

16. Shyshkina M. P. Formation and development of the cloud-based learning and research environment of higher educational institution: monograph / M. P. Shyshkina. – K. : UkrINTEI, 2015. – 256 p. (in Ukrainian)

17. Shyshkina M.P. Trends of development and standardization of requirements for cloud-based ICT learning tools / M.P.Shyshkina // Naukovyi visnyk Melitopolskoho derzhavnoho pedahohichnoho universytetu. Seriia: Pedahohika. – n.2(13). – 2014. – P.223-231. (in Ukrainian)

18. Shyshkina M. P. The formation and development of ICT tools of educational and research university environment within the concept of cloud computing / M. P. Shyshkina // Humanitarnyi visnyk DVNZ «Pereiaslav-Khmelnytskyi derzhavnyi pedahohichnyi universytet imeni Hryhoriia Skovorody». Tematychnyi vypusk «Vyshcha osvita Ukrainy u konteksti intehratsii do yevropeiskoho osvitnoho prostoru». – K. : Hnozys, 2014. – n. 5, vol. III (54). – App. 1. – P. 302–309. (in Ukrainian)

19. Shyshkina M. P. Models providing software access in the cloud-based learning and research environment / M. P. Shyshkina // Informatsiini tekhnolohii v osviti. – № 22. – 2015. – P. 120–129. (in Ukrainian)

20. Cusumano M. Cloud computing and SaaS as computing platforms / Michael Cusumano // Communications of the ACM. – 53(4). – 2010. – pp. 27-29. (in English)

21. Kendall M. Rank Correlation Methods / M. Kendall. – London: Charles Griffen & Company, 1948. (in English).

22. Kravtsov H. M. Methods and Technologies for the Quality Monitoring of Electronic Educational Resources [online] / H. M. Kravtsov // S. Batsakis et al. (eds.) Proc. 11-th Int. Conf. ICTERI 2015, Lviv, Ukraine, May 14-16. – 2015. – P. 311–325. – CEUR-WS.org/ Vol. 1356, ISSN 1613–0073, P. 311–325. – Available at: CEUR–WS.org/Vol–1356/paper_109.pdf (in English).

23. Mariya Shyshkina. The Hybrid Cloud-based Service Model of Learning Resources Access and its Evaluation [online] // Proceedings of the 12th International Conference on ICT in Education, Research and Industrial Applications. Integration, Harmonization and Knowledge Transfer / Ed. by Sotiris Batsakis, Heinrich C. Mayr, Vitaliy Yakovyna. - CEUR Workshop Proceedings. - vol.1356. - 2015. - pp.295-310. – Available from: http://ceur-ws.org/Vol-1614/paper_57.pdf (in English).

24. Mell P. The NIST Definition of Cloud Computing. Recommendations of the National Institute of Standards and Technology / P.Mell, T.Grance. - NIST Special Publication 800-145. NIST, Gaithersburg, MD 20899-8930, September 2011. (in English).

25. Shyshkina M. Emerging Technologies for Training of ICT-Skilled Educational Personnel / M. Shyshkina // Information and Communication Technologies in Education, Research, and Industrial Applications / V. Ermolayev, H. C. Mayr, M. Nikitchenko, A. Spivakovsky, G. Zholtkevych (Eds.). – Springer International Publishing, 2013. – P. 274–284. (in English).

26. Shyshkina M.P. Prospects of the Development of the Modern Educational Institutions' Learning and Research Environment: to the 15th Anniversary of the Institute of Information Technologies and Learning Tools of NAPS of Ukraine / M. P. Shyshkina, Y. G. Zaporozhchenko, H. M. Kravtsov // Information technologies in education. – Kherson. – 2014. – № 19. – P. 62–70. (in English).

27. Turner M. Turning software into a service / M. Turner, D. Budgen, P. Brereton // Computer. – 36 (10). – 2003. – pp. 38-44. (in English).

28. Vaquero L. M. EduCloud: PaaS versus IaaS cloud usage for an advanced computer science course / Vaquero Luis M. // IEEE Transactions on Education. – 54(4). - 2011. – pp. 590-598. (in English).

29. Wick D. Free and open-source software applications for mathematics and education / D. Wick // Proceedings of the twenty-first annual international conference on technology in collegiate mathematics. – 2009. – pp. 300-304. (in English).